The original text contains some typos and misprints that have been corrected in this version.

# Extending the topological analysis and seeking the real-space subsystems in non-Coulombic systems with homogeneous potential energy functions

Shant Shahbazian

*Faculty of Physics, Shahid Beheshti University, G. C., Evin, Tehran, Iran, 19839, P.O. Box 19395-4716.*

Tel/Fax: 98-21-22431661

E-mail: sh_shahbazian@sbu.ac.ir

[*] Corresponding author

**Abstract**
It is customary to conceive the interactions of all the constituents of a molecular system, i.e. electrons and nuclei, as Coulombic. However, in a more detailed analysis one may always find small but non-negligible non-Coulombic interactions in molecular systems originating from the finite size of nuclei, magnetic interactions, etc. While such small modifications of the Coulombic interactions do not seem to alter the nature of a molecular system in real world seriously, they are a serious obstacle for quantum chemical theories and methodologies which their formalism is strictly confined to the Coulombic interactions. Although the quantum theory of atoms in molecules (QTAIM) has been formulated originally for the Coulombic systems, some recent studies have demonstrated that most of its theoretical ingredients are not sensitive to the explicit form of the potential energy operator. However, the Coulombic interactions have been explicitly assumed in the mathematical procedure that is used to introduce the basin energy of an atom in a molecule. In this study, it is demonstrated that the mathematical procedure may be extended to encompass the set of the homogeneous potential energy functions thus relegating adherence to the Coulombic interactions to introduce the energy of a real-space subsystem. On the other hand, this extension opens the door for seeking novel real-space subsystems, apart from atoms in molecules, in non-Coulombic systems. These novel real-space subsystems, quite different from the atoms in molecules, call for an extended formalism that goes beyond the orthodox QTAIM. Accordingly, based on a previous proposal the new formalism, which is not confined to the Coulombic systems nor to the atoms in molecules as the sole real-space subsystems, is termed the quantum theory of real-space open subsystems (QTROS) and its potential applications are detailed. The harmonic trap model, containing non-interacting fermions or bosons, is considered as an example for the QTROS analysis. The QTROS analysis of bosonic systems is particularly quite unprecedented, not attempted before.

# 1. Introduction

The quantum theory of atoms in molecules (QTAIM) has gained a widespread recognition in the last twenty years in chemistry, molecular and solid-state physics, and even in molecular biology.[1-3] However, all applications of the QTAIM have been confined to the Coulombic systems namely, systems containing electrons and clamped nuclei interacting via the Coulombic potential. Even the recent extension of the QTAIM, termed the multi-component QTAIM (MC-QTAIM),[4-16] which goes beyond the clamped nucleus model and deals with the AIM analysis of certain types of non-Born-Oppenheimer molecular wavefunctions, is also confined to the Coulombic systems. Although it is understandable that the Coulombic systems are of prime interest in most applications in chemistry and physics, there are many non-Coulombic systems which are also interesting to be considered from viewpoint of the AIM analysis. However, before discussing examples of such systems, it must be emphasized that even for usual molecular systems the Coulombic interactions are just approximate potentials, albeit accurate enough for most practical applications, which are used usually in quantum chemical calculations. For highly accurate quantum description of an atomic or molecular system, various small but non-negligible non-Coulombic terms must be added to the Coulombic potential that weak internal magnetic interactions of electrons, originating from the L-S and the S-S couplings, and modifications originating from the finite size of nuclei are just examples. Accordingly, confining the QTAIM formalism to the Coulombic interactions is "artificial" and certainly against the basic idea that *atoms in molecules* are "real" objects emerging independent from the details of the models used to describe molecular systems.[17]

On the other hand, in recent decades a wealth of experimental and theoretical evidence has been accumulated demonstrating molecular-like structure for systems not traditionally considered as molecular systems. One may include in this list the "nuclear molecules" in nuclear physics,[18,19] various "exotic molecules" composed of fundamental particles other than electrons, protons and neutrons,[20-31] "artificial molecules" in condensed-matter physics,[32-35] and the "molecular Bose-Einstein condensates".[36-39] In considering such molecular-like systems the question emerges whether any underlying AIM structure is derivable from the wavefunctions of these systems. To answer this question one must apply the AIM analysis to these systems however, all such systems are intrinsically non-Coulombic in their nature and the formalism of the orthodox QTAIM must be modified to be applicable to these systems. Therefore, there is a real demand to extend the formalism of the orthodox QTAIM to non-Coulombic systems.

The programme of reconsidering the QTAIM formalism in the case of non-Columbic interactions was started sometime ago and it was demonstrated that the subsystem variational procedure and the subsystem hypervirial theorem are both insensitive to the nature of the potential energy operator as far as there is a bound quantum state in the system.[40,41] This is also true for the local zero-flux equation of the one-particle density, which is the equation of deriving the inter-atomic surfaces for both Coulombic and non-Coulombic systems.[40,41] However, upon considering the Hookean molecules, i.e. model systems where some of the Coulombic interactions have been replaced with the harmonic potential, it emerged that the AIM structures derived from the topological analysis were not the one expected based on "chemical intuition", which is routed in previous experiences with the Coulombic systems.[41] Thus, the use of topological analysis and the local zero-flux equation do not automatically guarantee that the emerging "real-space" subsystems are the usual AIM, also called topological atoms. In present study, more examples of exotic real-space subsystems in non-Coulombic systems are presented.

In contrast to the previous studies,[40,41] the focus of this contribution is on the part of the QTAIM formalism that is sensitive to the nature of the potential energy operator namely, the basin energy of an atom in a molecule.[1] Accordingly, the definition of the basin energy is extended beyond the Coulombic potential energy function demonstrating that for the subset of homogeneous potential energy functions the regional virial theorem may be used to derive well-defined, origin-independent, basin energies.

## 2. The generalized subsystem virial theorem for the homogeneous potential energy functions

The atomic/regional theorems, emerging from the subsystem hypervirial theorem,[9,42,43] are insensitive to details of the potential energy operator and are true as far as a system is composed of a single type of quantum particles and there is a bound stationary state emerging from the interaction of quantum particles with each other and the external fields. This insensitivity is compelling since the orthodox formalism may be employed with least modifications for non-Coulombic systems however, the regional/basin energies have been derived employing explicitly the properties of the Coulombic potential (see particularly section 6.3 in [1]). In the present section, the very definition of the basin energy is extended to include the set of the homogeneous potential energy functions (for an elementary discussion on the homogeneous potential energy functions see chapter 14 in [44]).

A homogeneous potential energy function for a typical *N*-particle system has the following property: $\hat{V}(s\vec{r}_1,...,s\vec{r}_N) = s^n \hat{V}(\vec{r}_1,...,\vec{r}_N)$, where *s* is an arbitrary scaling parameter and *n* is the degree of homogeneity.[44] It is straightforward to demonstrate that for this set of potential energy functions the following relation holds: $\hat{V}(\vec{r}_1,...,\vec{r}_N) = (1/n)\sum_{k=1}^{N} \vec{r}_k \cdot \vec{\nabla}_k \hat{V}(\vec{r}_1,...,\vec{r}_N)$,[44] where $\vec{r}_k$ are the vectors describing the position of each of the *N* particles; the Coulombic potential is a special member of this set where $n = -1$.[1] It is evident that $\sum_{k=1}^{N} \vec{r}_k \cdot \vec{\nabla}_k$ is a projection operator and it is called the virial operator. It is also straightforward to demonstrate that the virial theorem holds generally for any stationary state of an *N*-particle system: $2\langle \hat{T} \rangle = \langle \sum_{k=1}^{N} \vec{r}_k \cdot \vec{\nabla}_k \hat{V} \rangle$,[44] where $\hat{T}$ is the sum of the kinetic energy operators of all quantum particles, $\hat{T} = \sum_{k=1}^{N} \hat{t}_k = (-\hbar^2/2m)\sum_{k=1}^{N} \nabla_k^2$, while $\langle ... \rangle$ is used to denote the mean value of the operators for a stationary state. For systems where the potential energy operator is a homogeneous function the virial theorem simplifies to: $2\langle \hat{T} \rangle = n\langle \hat{V} \rangle$.[44]

The local form of the virial theorem derived from the subsystem hypervirial theorem is as follows:[1]

$$2T(\vec{q}) = -V^T(\vec{q}) + L(\vec{q}) \qquad (1)$$

In this equation $T(\vec{q})$ is the kinetic energy density introduced as:

$T(\vec{q}) = \int d\tau' \, \Psi^* \left( \sum_{k=1}^{N} \hat{t}_k \right) \Psi = N \int d\tau' \, \Psi^* \hat{t}_q \Psi = -(1/2)tr[\vec{\sigma}(\vec{q})] + (1/2)L(\vec{q})$, where

the second equality originates from the indistinguishability of quantum particles. $V^T(\vec{q}) = -\vec{q} \cdot (\vec{\nabla} \bullet \vec{\sigma}(\vec{q})) + \vec{\nabla} \cdot (\vec{q} \bullet \vec{\sigma}(\vec{q}))$ is the total virial density (the symbol $\bullet$ is used to emphasize the dyadic nature of the product) while $L(\vec{q}) = (-\hbar^2/4m)\nabla^2 \rho(\vec{q})$ where $\rho(\vec{q}) = N\int d\tau' \Psi^*\Psi$ is the one-particle density of quantum particles ($d\tau'$ implies summing over spin variables of all quantum particles and integrating over spatial coordinates of all quantum particles except a typical particle denoted by $\vec{q}$). The stress tensor density is the key density that both kinetic and total virial densities are based on while the Schrödinger-Pauli-Epstein variant is used in this study:

$$\vec{\sigma}(\vec{q}) = \left(\frac{N\hbar^2}{4m}\right)\int d\tau' \{\Psi^*(\vec{\nabla}\vec{\nabla}\Psi) + \Psi(\vec{\nabla}\vec{\nabla}\Psi^*) - (\vec{\nabla}\Psi^*)(\vec{\nabla}\Psi) - (\vec{\nabla}\Psi)(\vec{\nabla}\Psi^*)\}.^{[1]}$$

It is timely to emphasize that stress tensor density is not unique and the Schrödinger-Pauli-Epstein variant is just one member of the infinitely large family of the stress tensor densities.[45] For a real-space subsystem, e. g. AIM, enclosed by the zero-flux surfaces, $\Omega$, based on Gauss's theorem one derives: $L(\Omega) = (-\hbar^2/4m)\int_\Omega d\vec{q}\ \nabla^2 \rho(\vec{q}) = (-\hbar^2/4m)\oint_{\partial\Omega} dS\ \vec{\nabla}\rho(\vec{q}) \cdot \vec{n}(\vec{q}) = 0$ ($\vec{n}(\vec{q})$ is the unit vector orthogonal to the zero-flux surface). Also, $T(\Omega) = \int_\Omega d\vec{q}\ T(\vec{q})$ and $V^T(\Omega) = \int_\Omega d\vec{q}\ V^T(\vec{q})$ are basin kinetic and total virial energies, respectively, and the regional/subsystem virial theorem is as follows:[1]

$$2T(\Omega) = -V^T(\Omega) \qquad (2)$$

It is important to realize that the total virial density is composed of two contribution, one originating directly from the virial operator and called basin virial density:

$$V^B(\vec{q}) = \int d\tau' \, \Psi^*\left(-\sum_{k=1}^{N}\vec{r}_k \cdot \vec{\nabla}_k \hat{V}\right)\Psi = N\int d\tau' \, \Psi^*\left(-\vec{q} \cdot \vec{\nabla}_q \hat{V}\right)\Psi = -\vec{q} \cdot \left(\vec{\nabla}_q \bullet \vec{\sigma}(\vec{q})\right)$$

and another term originating from the assumed zero-flux surfaces as boundaries of subsystems and called surface virial density: $V^S(\vec{q}) = \vec{\nabla}.(\vec{q}\bullet\vec{\sigma}(\vec{q}))$. It is straightforward to demonstrate that the surface virial is null for the total system and this fact differentiates the virial theorem of total system with that of the real-space subsystems.[1]

At the mechanical equilibrium,[1] the Hamiltonian of an N-particle system with a homogeneous potential energy is:

$$\hat{H} = \hat{T} + \hat{V} = \sum_{k=1}^{N}\left(\hat{t}_k + (1/n)\vec{r}_k \cdot \vec{\nabla}_k \hat{V}\right) = \sum_{k=1}^{N}\hat{h}_k \qquad (3)$$

Based on this equation the energy density is:

$$E(\vec{q}) = \int d\tau' \, \Psi^*\left(\sum_{k=1}^{N}\hat{h}_k\right)\Psi = N\int d\tau' \, \Psi^*\hat{h}_q \Psi$$

$$= N\int d\tau' \, \Psi^*\left(-(\hbar^2/2m)\nabla_q^2 + (1/n)\vec{q}\cdot\vec{\nabla}_q\hat{V}\right)\Psi = T(\vec{q}) - (1/n)V^B(\vec{q}) \qquad (4)$$

Integration of the energy density in the whole space ($R^3$) yields the total energy of the system: $E = \langle\hat{T}\rangle + \langle\hat{V}\rangle$, while based on the virial theorem for total system one derives: $E = (1+2/n)\langle\hat{T}\rangle = (1+n/2)\langle\hat{V}\rangle$. However it is well-known if the integration is done on a real-space subsystem ($\Omega \subset R^3$), then the resulting basin energy, because of the origin-dependence of the basin virial density, is also origin dependent, which is plainly an unpleased feature.[1] To overcome this problem, inspired by the regional virial theorem, equation (2), the following modified energy density and basin energy are introduced:

$$E(\vec{q}) = T(\vec{q}) - (1/n)V^T(\vec{q}) = T(\vec{q}) - (1/n)(V^B(\vec{q}) + V^S(\vec{q}))$$

$$E(\Omega) = \int_\Omega d\vec{q}\ E(\vec{q}) = T(\Omega) - (1/n)V^T(\Omega) \tag{5}$$

Using equation (2) as the regional virial theorem the basin energy may be expressed just by the regional kinetic or total virial energies:

$$E(\Omega) = (1 + 2/n)T(\Omega) = -(1/2 + 1/n)V^T(\Omega) \tag{6}$$

For the special case of the Coulombic potentials equation (6) recovers the well-known results derived from the orthodox formalism: $E(\Omega) = -T(\Omega) = (1/2)\ V^T(\Omega)$.[1]

For $N$-particle systems with one- and two-particle interactions the potential energy operator is: $\hat{V}(\vec{r}_1,...,\vec{r}_N) = \sum_{k=1}^{N} \hat{v}_k(\vec{r}_k) + \sum_{i>j}^{N} \hat{v}_{ij}(\vec{r}_i, \vec{r}_j)$. The role of the virial operator is the projection of the two-particle terms into "pseudo" one-particle contributions and this is easily seen for a two-particle system: $\hat{v}_{12} = (1/n)(\vec{r}_1 \cdot \vec{\nabla}_1 \hat{v}_{12} + \vec{r}_2 \cdot \vec{\nabla}_2 \hat{v}_{12})$; these "pseudo" one-particle contributions make it possible to introduce the virial density, bypassing the need to introduce potential energy density explicitly.[1] For the subset of $N$-particle systems without two-particle interactions, i.e. non-interacting systems trapped in external potentials, the relation between one-particle interactions and the virial operator is as follows: $\hat{v}_k = (1/n)\vec{r}_k \cdot \vec{\nabla}_k \hat{v}_k$. Accordingly, one may now introduce the potential energy density directly: $V(\vec{q}) = \int d\tau'\ \Psi^* \left( \sum_{k=1}^{N} \hat{v}_k \right) \Psi = N \int d\tau'\ \Psi^* \hat{v}_q \Psi$, which is equal to the basin virial density. The local and regional forms of the virial theorem are then transformed as follows:

$$2T(\vec{q}) = nV(\vec{q}) - V^s(\vec{q}) + L(\vec{q})$$

$$2T(\Omega) = nV(\Omega) - V^s(\Omega) \tag{7}$$

The energy density and basin energies for the real-space subsystems is then introduced as follows:

$$E(\vec{q}) = T(\vec{q}) + V(\vec{q}) - (1/n)V^s(\vec{q})$$

$$E(\Omega) = \int_\Omega d\vec{q}\; E(\vec{q}) = T(\Omega) + V(\Omega) - (1/n)V^s(\Omega)$$

$$= (1 + 2/n)T(\Omega) = (1 + n/2)V(\Omega) - (1/2 + 1/n)V^s(\Omega) \tag{8}$$

These equations vividly demonstrate that apart from the potential energy density originating from the interaction of each quantum particle with the external field, the surface virial also contributes to the basin energy. Assuming $\Omega = R^3$ the surface virial vanishes and the equations are indistinguishable from those derived for the total system independently.

## 3. The topological analysis of non-Coulombic systems: The harmonic trap model

The topological analysis of the one-particle density yields the topological structure, through identifying critical points (CPs) and the boundaries between real-space subsystems. Particularly, the local zero-flux equation, $\vec{\nabla}\rho(\vec{q}) \cdot \vec{n}(\vec{q}) = 0$, is used to determine the zero-flux surfaces that act as inter-atomic boundaries.[1] However, these surfaces are just a small subset of the zero-flux surfaces emerging from the equation.[40,46,47] It has been demonstrated that the zero-flux surfaces that are not acting as the boundaries of topological atoms may found interesting applications; the "morphologies" of the real-space subsystems which they are shaping are different from topological atoms, [48-55] and even more exotic (from the viewpoint of their morphology) real-space subsystems emerge from the net zero-flux equation, $\int_\Omega d\vec{q}\ \nabla^2 \rho(\vec{q}) = 0$, as demonstrated recently.[47,56] All these studies point to the fact that even for the Coulombic systems the topological analysis may yield a wide spectrum of real-space subsystems apart from the topological atoms. Accordingly, it is tempting to consider what kind of real-space subsystems may emerge from the topological analysis of non-Coulombic systems. In the rest of this section the harmonic trap model is considered for this purpose.

The model of $N$ quantum particles confined in a harmonic trap has been widely used to model the Bose-Einstein condensation in trapped dilute gases,[57-68] and more recently in trapped Fermi gases.[69-73] A simplified model of trap may be constructed assuming a non-interacting system of quantum particles in an external isotropic harmonic trap, as a homogeneous potential, $n = 2$, with the following Hamiltonian: $\hat{H} = \sum_{k=1}^{N} \hat{h}_k = \left(-\hbar^2/2m\right)\sum_{k=1}^{N} \{\nabla_k^2 - \alpha^2(x_k^2 + y_k^2 + z_k^2)\}$, where $\alpha = 2\pi f m/\hbar$ and $f$ is the frequency of mechanical vibration of the particle in the trap.[44] The spectrum of the eigenfunctions and eigenvalues of the one-particle Hamiltonian, $\hat{h}_K \phi_{v_1 v_2 v_3} = \varepsilon_{v_1 v_2 v_3} \phi_{v_1 v_2 v_3}$, is well-known ($v_1, v_2, v_3$ are the quantum numbers),[44] e.g. $\phi_{000}(x,y,z) = (\alpha/\pi)^{3/4} Exp[-(\alpha/2)(x^2 + y^2 + z^2)]$, $\varepsilon_{000} = 3\pi\hbar f$ and $\phi_{100}(x,y,z) = (4\alpha^5/\pi^3)^{1/4} x Exp[-(\alpha/2)(x^2 + y^2 + z^2)]$, $\varepsilon_{100} = 5\pi\hbar f$. The wavefunction of the system may be constructed based on the statistics of the trapped particles. In the ground state of the system filled with non-interacting bosons all particles are at the lowest one-particle energy state, $E_0^{Boson} = N\varepsilon_{000} = 3N\pi\hbar f$, and neglecting the spin variable, the spatial part of the wavefunction is a simple product of the one-particle eigenfunctions associated to the lowest one-particle energy state:

$$\Psi_{Boson} = \prod_{k=1}^{N} \phi_{000}(x_k, y_k, z_k) = (\alpha/\pi)^{3N/4} Exp\left[-(\alpha/2)\sum_{k=1}^{N}(x_k^2 + y_k^2 + z_k^2)\right].$$

On the other hand, if the trap is filled with fermions then the spin variable is of pivotal importance and the spin-eigenfunctions, instead of the spatial eigenfunctions, must be used to construct the fermionic wavefunction, $\begin{cases}\psi_{v_1v_2v_3} = \phi_{v_1v_2v_3}\alpha \\ \overline{\psi}_{v_1v_2v_3} = \phi_{v_1v_2v_3}\beta\end{cases}$ ($\alpha$ and $\beta$ are the spin eigenfunctions). The Pauli Exclusion Principle dictates a $N \times N$ determinant, composed of the spin-eigenfunctions, as the ground state wavefunction of the system:

$$\Psi_{Fermion} = (N!)^{-1/2}\sum_{i=1}^{N!}(-1)^{p_i}\hat{P}_i\left[\psi_{000}(x_1,y_1,z_1)\,\overline{\psi}_{000}(x_2,y_2,z_2)\,\psi_{100}(x_3,y_3,z_3)...\right],$$

where $\hat{P}_i$ is the permutation operator generating all possible permutations of particles within the spin-eigenfunctions while $p_i$ is the number of transpositions/exchanges (the wavefunction is a linear combination of such determinants if the determinants are describing degenerate ground states).[44] The ground state energy of the fermionic system is: $E_0^{Fermion} = 3N\pi\hbar f + 2\pi\hbar f \sum_{v_1}\sum_{v_2}\sum_{v_3} n_{v_1v_2v_3}(v_1 + v_2 + v_3)$, where $n_{v_1v_2v_3}$ is the occupation number of the one-particle energy states denoted by the quantum numbers $v_1, v_2, v_3$ and is always equal to two, one or zero.

The formalism of the QTAIM is insensitive to the statistics of quantum particles however, according to the best of author's knowledge, no previous QTAIM analysis of a bosonic system has been done. This is understandable since only many-electron systems have been considered within the context of the QTAIM.[1] The one-particle density and its gradient vector field for the bosonic system are as follows:

$$\rho_{Boson}(x, y, z) = N(\alpha/\pi)^{3/2} Exp[-\alpha(x^2 + y^2 + z^2)]$$

$$\vec{\nabla}\rho_{Boson} = -N\sqrt{4\alpha^5/\pi^3} Exp[-\alpha r^2] \vec{r} \qquad (9)$$

Since the one-particle density is isotopic, the gradient vector field is written in the spherical polar coordinate system ($r, \theta, \varphi$): $r = \sqrt{x^2 + y^2 + z^2}$ and $\vec{r} = r\vec{r}_0$, where $\vec{r}_0 = \vec{i} \sin\theta\cos\varphi + \vec{j} \sin\theta\sin\varphi + \vec{k}\cos\theta$ is the unit vector.[44] It is evident from these equations that the topological structure of the gradient vector field is independent from the number of particles and from the equation: $\vec{\nabla}\rho_{Boson} = 0$, just a single (3, -3) CP emerges at the origin of the coordinate system. The one-particle density monotonically decays from its maximum value at the origin $\rho_{Boson}(0,0,0) = N(\alpha/\pi)^{3/2}$ and this pattern is reminiscent of the one-electron density of atoms.[1] This similarity is suggestive that the ground state of the bosonic aggregate, trapped in the external harmonic potential, independent from the number of trapped bosons, is similar to a single atom (a "giant atom" if $N \to \infty$). Interestingly, this is also in line with the description of the Bose-Einstein condensate at its ground state as a "super-atom".[57] Evidently, just a single topological atom emerges from the topological analysis and the zero-flux surfaces emerging from the local zero-flux equation are all crossing the CP. In the case of the fermionic system the explicit form of the ground state one-particle density depends on the number of particles and only two cases, $N = 2, 8$, are considered here. For a two-particle system the one-particle density and its gradient vector field for the system are as follows:

$$\rho_{Fermion}^{N=2}(r) = \sqrt{4\alpha^3/\pi^3}\, Exp[-\alpha r^2]$$

$$\vec{\nabla}\rho_{Fermion}^{N=2} = -\sqrt{16\alpha^5/\pi^3}\, Exp[-\alpha r^2]\, \vec{r} \qquad (10)$$

These equations clearly demonstrate that the two-particle fermionic system is quite similar to the bosonic system and the structure of a single atom emerges from the topological analysis. For the eight-particle system, $N=8$, the Pauli Exclusion Principle dictates the occupation of the three degenerate one-particle lowest energy excited states $\phi_{100}, \phi_{010}, \phi_{001}$ (a "closed-shell" configuration), apart from the ground one-particle $\phi_{000}$ state which is also occupied for $N=2$ system. The one-particle density and its gradient vector field for this system are as follows:

$$\rho_{Fermion}^{N=8}(r) = \sqrt{4\alpha^3/\pi^3}\,(1+2\alpha r^2)Exp[-\alpha r^2]$$

$$\vec{\nabla}\rho_{Fermion}^{N=8}(r) = \sqrt{16\alpha^5/\pi^3}\,(1-2\alpha r^2)Exp[-\alpha r^2]\, \vec{r} \qquad (11)$$

In contrast to the equations (9) and (10), the one-particle density is not monotonically decaying in this system and from the equation: $\vec{\nabla}\rho_{Fermion}^{N=8} = 0$, two kinds of CPs emerge. A CP is located at the center of the coordinate system and infinite numbers of CPs are all located on a spherical surface around the center of the coordinate system with the radius: $r_{CP} = 1/\sqrt{2\alpha}$. The amount of one-particle density at the central CP is: $\rho_{Fermion}^{N=8}(0) = \sqrt{4\alpha^3/\pi^3}$, while on the spherical surface one finds: $\rho_{Fermion}^{N=8}(1/\sqrt{2\alpha}) = 2Exp[-1/2]\sqrt{4\alpha^3/\pi^3}$. Evidently, the amount of one-particle density is larger at the spherical shell and the central CP is a global minimum or a (3, +3) CP whereas the CPs on the spherical surface are "non-isolated" (1, -1) CPs that have been rarely observed in molecular systems.[74] Instead of the well-known "point" attractors with rank 3, e.g. (3, -3) or (3, -1), in this system one is faced with a "global" attractor with rank 1, i.e. (1, -1), which is a spherical surface with infinite numbers of degenerate point attractors; a similar global attractor in the one-electron density of the 2S excited state of hydrogen atom also appears.[75] Based on the emerging topological structure, this system also seems to be composed of a single real-space subsystem though it is not a topological atom. Finally, one infers from the comparison of the eight-particle bosonic and fermionic systems that statistics of particles has a pivotal role on the topological structure which does not seem to be noticed previously.

## 4. Conclusion and prospects

The programme of extending the QTAIM formalism to non-Coulombic systems widens the applications of the theory and in this regard, it is similar to the ongoing programme of extending the QTAIM to the multi-component systems. Sometime ago it was proposed that the real-space subsystems emerging from topological analysis do not need to be similar to the topological atoms and a generalized framework called Quantum Theory of Real-space Open Subsystems (QTROS) was developed to deal with all types, rather than just the topological atoms, of real-space subsystems.[47] While in that paper only the Coulombic systems were conceived as targets, the present contribution demonstrates that the QTROS may be conceived as a general theory that deals with both the Coulombic and non-Coulombic systems composed of a single type of quantum particles interacting with each other and external fields through the homogeneous potentials. Apart from the previously considered examples,[41] and those considered in this chapter, a large number of interesting systems, some indicated in the first section, remain to be considered within context of the QTROS. However, a completely comprehensive theory must encompass also quantum systems containing particles that their interaction potentials are inhomogeneous functions. This is also important in extending the QTAIM analysis further since upon adding new, albeit small, terms to the Coulombic potentials the resulting potential energy operator is inevitably inhomogeneous.

The comparative analysis of the real-space subsystems emerging in fermionic and bosonic systems is another novel aspect of the present study. This is an interesting area for future studies since it may reveal the "local" role of Pauli Exclusion Principle in molecular systems. Pauli "repulsions" and associated steric interactions are usually invoked in both qualitative and quantitative analysis to rationalize conformational selections, tracing molecular stresses and instabilities. However, most of such analyzes are based on indirect methods and one may hope that a direct comparative QTAIM analysis on a fermionic system and associated bosonic counterpart may reveal a more detailed picture of the role of the statistics on the local interactions in molecular systems.

## Acknowledgments

The author is grateful to Masume Gharabaghi and Ángel Martín-Pendás for their detailed reading of a previous draft of this paper and helpful suggestions.